**Intrinsic Ductility from Shear Amorphization: From Pure Metals to Multi-Principal-Element Alloys**

Morgan R. Jones[1], Duane D. Johnson[2,3], and N. Argibay[4,*]

[1]Center for Computing Research, Sandia National Laboratories, Albuquerque, NM, USA
[2]Division of Materials Science and Engineering, Ames National Laboratory, IA, USA
[3]Materials Science and Engineering, Iowa State University, Ames, IA, USA
[4]The Scientific Review, MT, USA

[*]Corresponding author: nargibay@gmail.com

*Direct links between electronic structure and intrinsic ductility remain elusive for metals. A framework is proposed that reduces the complexities of valence charge distribution, band filling, and shear strain effects into structure-property relationships describing the intrinsic ductility of metals and alloys. Rather than relying on crystal cleavage and dislocation nucleation at preexisting crack tips, we show that a lower energy fracture criterion, i.e., the activation energy density for amorphization, enables accurate predictions of both intrinsic ductility and ductile-to-brittle transition temperatures. From analytical expressions and tabulated ab-initio stiffness constants, lattice parameters, and binary interaction energies, we present a unified theory that reconciles ductile flow in pure metals and solid-solution alloys. Phase diagrams generated for the Nb-Ta-V-Ti system simultaneously explain its high strength and room-temperature tensile ductility, validating this framework as a practical one for rapid design of structural multi-principal-element alloys.*

## Introduction

The design of structural alloys with a combination of high strength and intrinsic ductility is critical to addressing challenges in energy sustainability. Alloying seeks to combine dissimilar elements to introduce strength while maintaining ductility. There are a multitude of mechanisms with a firm theoretical foundation for the strengthening of metals, *e.g.*, grain refinement[1–3], work-hardening[4,5], solution effects[6–11], precipitate-particle inclusions[12–14], and thermodynamic stabilization of defects like grain boundaries[15,16], yet the prediction of ductility remains elusive. The history and evolution of ductility models spans the seminal empirical work of Pugh[17], who first identified the ratio of shear and bulk moduli ($G/B$) as an indicator[18] of whether materials would favor crystal slip over cleavage, to Pettifor[19,20], who suggested that a positive Cauchy pressure favors ductility (*i.e.*, $C_{12}$ - $C_{44}$ > 0 for cubic crystals, from stiffness constants). Rice[21] introduced the concept of unstable stacking fault energy ($\gamma_{usf}$) to quantify the potential-energy barrier for dislocation slip, and proposed a ductility model, now referred to as the D-parameter, which is the ratio of $\gamma_{usf}$ and the energy barrier for cleavage via the Griffith criterion[22,23]. The latter relies on the surface energy ($\gamma_{surf}$), resulting in the expression for the D-parameter, $D \propto \gamma_{usf}/\gamma_{surf}$, which, as we show, requires far larger (with inconsistent proportionality) activation energy than that for nucleating an amorphous interface. The D-parameter is predictive of whether cleavage or dislocation emission are energetically favorable; however, Rice pointed out that the accuracy of this approach is limited by the complexity of the multi-modal fracture stress state near a crack tip, and also necessitates corrections for crystal orientation-dependent slip. Despite this important limitation, the D-parameter has been useful as a screening tool for new alloy compositions, particularly in the enormous composition space encompassed by concentrated solid solutions, including both multi-principal-element alloys (MPEAs) and high-entropy alloys (HEAs)[24–30].

MPEAs have emerged as a focus area for overcoming historical limitations of refractory metals and their dilute alloys[31–34]. Until the seminal work of Cantor[35] and Yeh[36], alloying was largely confined to the use of a primary element as a majority species, as with Fe in steels, Ni and Co in superalloys for turbomachinery, Al and Ti for automotive and aerospace applications, specialized steels used in fission reactor containment, and W-rich alloys needed for prolonged plasma containment in planned commercial fusion reactors. Concentrated solid solutions with three or greater combinations of elements in non-dilute concentrations have been established as a route for improving strength, ductility, and thermal (phase) stability at extremes of temperature, stress, and strain rate[32,35–40]. Notably, concentrated combinations of refractory metals like W, Mo, Nb, Ta, and V, have been identified as a pathway for addressing a central challenge to enabling next-generation energy production and propulsion systems, i.e., imparting and retaining low-temperature ductility in high-strength alloys[40–42].

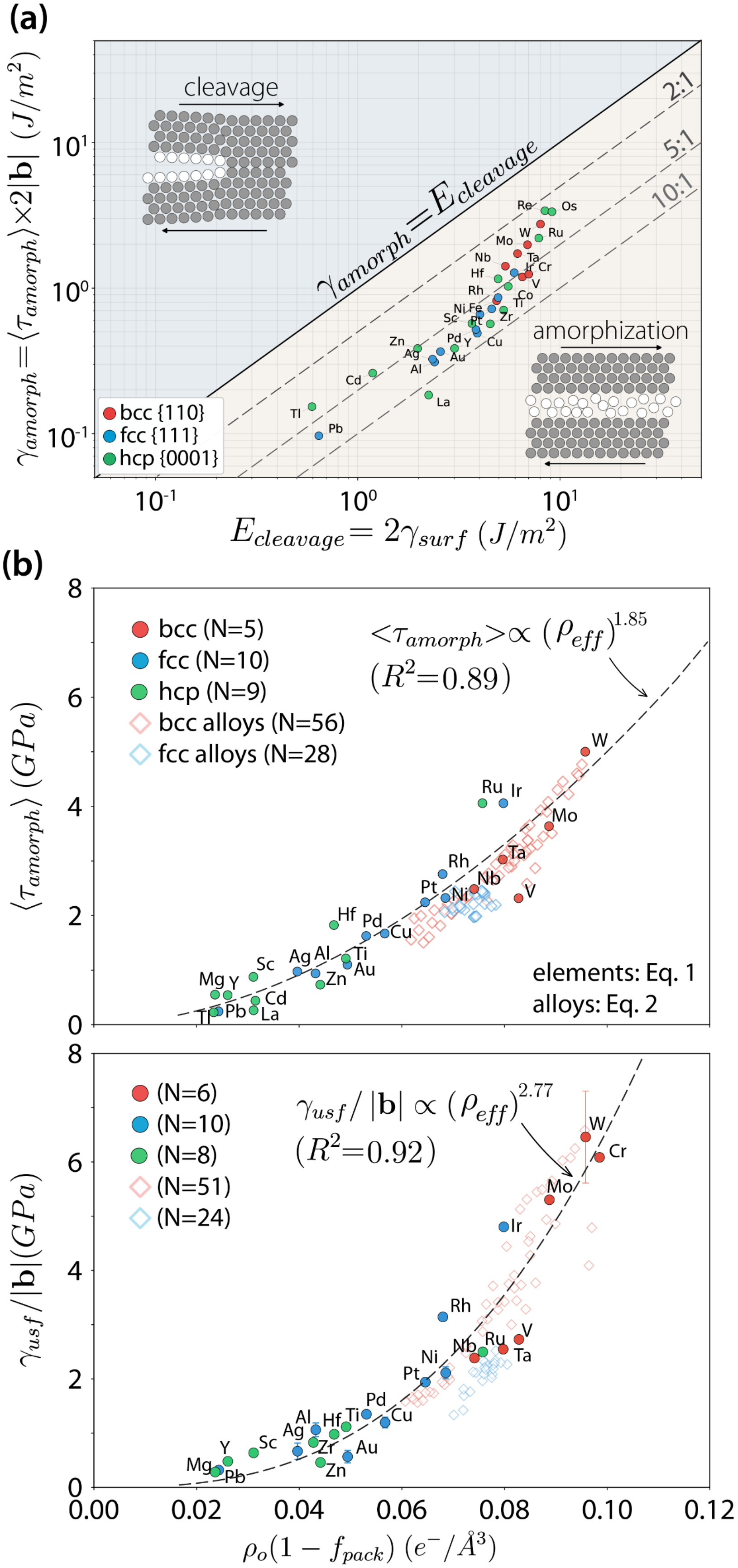

***Figure 1:*** *(a) Comparison of the activation energy for cleavage via crack propagation*[21] *to the activation energy for generation of an amorphous interface*[43]*, analogous to a high-energy grain boundary*[44]*, showing that the latter is the energetically preferred mechanism. (b) Interstitial charge density correlations to the activation stresses for amorphization (top panel) and slip (bottom panel), for both elements and alloys.*

From an electronics-structure and energetics perspective, higher strength is achieved by increasing the activation energy for deformation mechanisms associated with crystal slip. However, ductility is the ability to deform without fracture, which is a competition between the activation energy for the most favorable slip mechanism and the activation energy for fracture. Historically, as with the seminal work by Rice, the fracture mechanism has been assumed to be proportional to surface energy, i.e., the energy associated with crack propagation, and by extension, crystal cleavage[21]. Here, we examine the use of the more energetically favorable activation energy for the formation of an amorphous interface as the limiting fracture criterion[43,44], as described in **Figure 1**. We also propose refinements to the prediction of slip activation for cubic crystal systems based on interstitial-charge density in metals[45], as summarized in **Figure 2**. Limiting mechanisms for both slip and fracture are defined on an energy-density basis, and their ratio is shown to accurately predict intrinsic ductility in both pure metals and alloys. The proposed structure-property relationships linking interstitial-charge density, slip activation energies, and amorphization activation energy enable predictions of intrinsic ductility in random concentrated solid-solution alloys. This framework is exclusively reliant on physically-meaningful parameters, including element pair-interaction energies[46] and rule-of-mixtures combinations of crystal structure-matched lattice constants[47] and interstitial-charge densities[45].

## Intrinsic strength and ductility from charge density

### *Amorphization, not cleavage, as the limiting strength mechanism*

As shown in **Figure 1a**, the energy density for cleavage ($2\gamma_{surf}$) is 3x-10x larger than the energy density ($\gamma_{amorph}$) for the shear-activated formation of an amorphous interface [21], analogous to a high-energy grain boundary[44] with structural disorder comparable to a liquid structure and exhibiting vanishingly low strength[43]. A central premise of the proposed approach for prediction of intrinsic ductility is that the stresses and energy densities associated with crack propagation or cleavage are significantly higher than those for the formation of an amorphous interface. It is only necessary to overcome the lower amorphization energy barrier to nucleate fracture, as it follows that amorphous shear bands[48] will form and meet a free surface. For example, the intersection of an amorphous shear band and an internal defect, like a void, sequentially leads to cleavage by reducing the localized energy required for cleavage. The energy density for amorphization of pure metals and alloys is given in **Equations 1 and 2**, based on derivations in earlier work[43,44]. For elements, the temperature-dependent activation stress for amorphization based on energy density, $\langle\tau_{amorph}\rangle$, is the product of enthalpy of fusion, $H_f$, liquid density at melting temperature, $\rho_{liq}$, molar mass, $M$, and melting temperature, $T_m$. The ratio of liquid density and molar mass is the average atomic volume of the structurally amorphous solid state. For alloys, a rule-of-mixtures amorphization stress is calculated based on elemental atomic concentrations, $x_i$, and an additional mixing free energy term introduced to correct deviation from ideal behavior, as with the regular solution model.

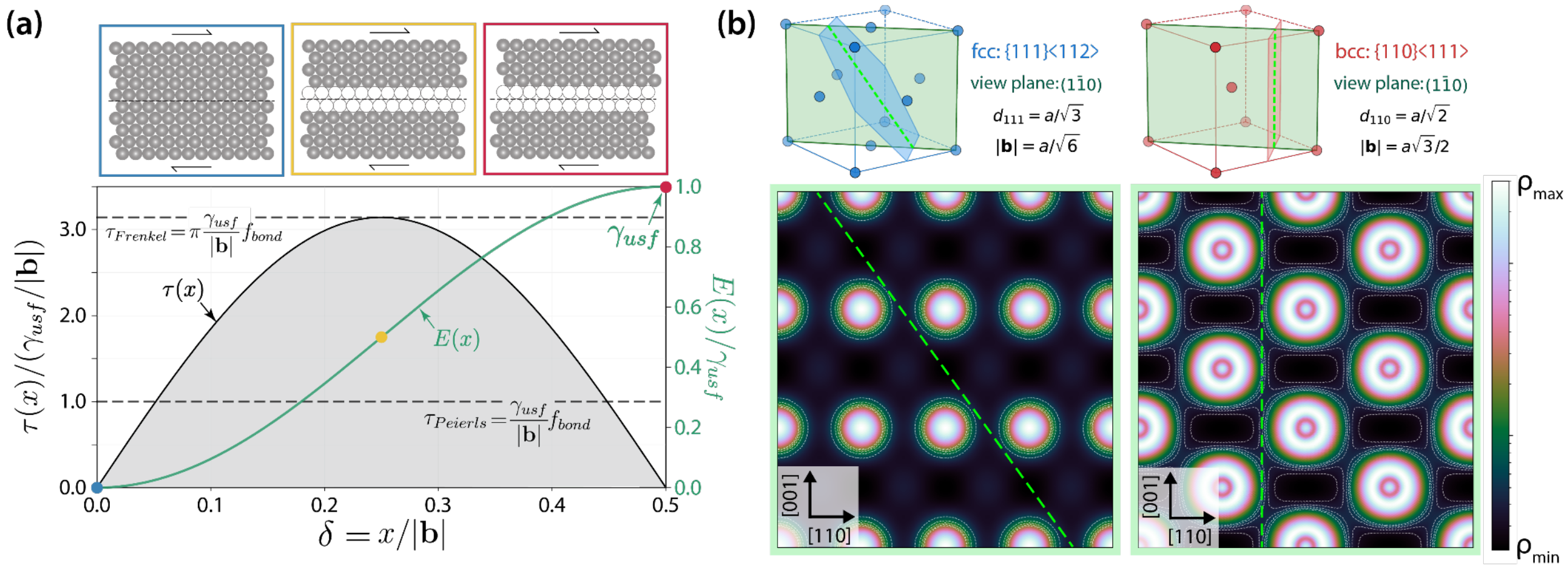

***Figure 2:*** *(a) Diagram showing the generalized stacking fault energy curve and its relationship to the activation energy densities of the two limiting barriers to slip. (b) Illustration of the preferred slip systems for cubic crystals and cross-sectional images showing DFT-based interstitial charge density distributions orthogonal to the preferred slip planes (which happens along the dotted green lines) for two prototypical cases: Cu, which exhibits strong stiffening under shear deformation, and W, which exhibits effectively no stiffening during shear.*

For the ensuing calculations, the enthalpy of mixing for random concentrated solid-solutions is determined from tabulated DFT-based binary interaction energies from Chen et al.[46] for cubic crystal structures. Following Vegard's law (see supplemental), alloy atomic volumes, $V_{alloy}$, are calculated at absolute zero temperature based on the equilibrium DFT lattice constants provided by Johnson *et al.*[45], and a 10% reduction in volume applied as an approximation (Lindemann criterion[49]) for the difference between crystalline and solid-state amorphous volumes, $V_{liq,alloy}$ [44]. As the amorphization stress captures the free energy associated with a stress-activated change in structural ordering, the mixing free-energy term includes a normalization by the change in coordination number from ideal crystals ($z_{crys}$ = 8 and 12 for BCC and FCC, respectively) and liquids, estimated for metals[50,51] to be $z_{liq}$ = 10.5 ± 1.2. Both corrections, for the volume expansion and bond-density change based on coordination numbers, only amount to a change of ~20% for BCC and ~10% for FCC from ideal values.

**elements**:
$$\langle \tau_{amorph} \rangle = H_f \frac{\rho_{liq}}{M} \left(1 - \frac{T}{T_m}\right) \quad \textbf{Eq. 1}$$

**alloys**:
$$\langle \tau_{amorph} \rangle = \sum_i x_i \langle \tau_{amorph,i} \rangle + \left(1 - \frac{z_{crys}}{z_{liq}}\right) \frac{\Delta G_{mix}}{V_{liq,alloy}} \left(1 - \frac{T}{T_{m,alloy}}\right) \quad \textbf{Eq. 2}$$

For alloys, it is possible to calculate all necessary parameters for using **Equations 3-5** [43].

$$V_{liq,alloy} = 0.9 \sum_i V_{alloy,i} \quad \textbf{Eq. 3}$$

$$\Delta G_{mix} = H_{mix} - T\Delta S_{mix} = H_{mix} + RT \sum_i x_i ln\left(x_i\right) \quad \textbf{Eq. 4}$$

$$T_{m,alloy} = \sum_i x_i T_{m,i} + \left(1 - \frac{z_{crys}}{z_{liq}}\right) H_{mix} \sum_i \frac{T_{m,i}}{x_i H_{f,i}} \quad \textbf{Eq. 5}$$

The energy density to achieve amorphization was shown[44] to agree with those of high-energy grain boundaries, and the associated stress, $\gamma_{amorph} = \langle \tau_{amorph} \rangle \cdot \delta$, where the thickness of the amorphous interface or grain boundary is approximately twice the magnitude of a Burgers vector, $\delta \approx 2\left|b\right|$. **Figure 1b** shows that the limiting strength $\langle \tau_{amorph} \rangle$ can be determined directly from DFT-calculated interstitial charge density ($\rho_o$) and crystal structure packing fraction ($f_p$), . , where $\rho_o$ is the ratio of *ab initio* interstitial electron count (IEC) and lattice constants for elements and rule-of-mixtures (of lattice-matched) values for random concentrated solid-solution alloys. The effective (structure-corrected) interstitial charge density, $\rho_{eff} = \rho_o\left(1 - f_{pack}\right)$, where $f_{pack}$ is 0.7404 for FCC/HCP and 0.6802 for BCC crystals. The maximum strength of metals defined by **Equation 1**, which is an intrinsic value when calculated at absolute zero temperature, is an expression that uses no fitting parameters. **Equations 2-5** enable predictions of the fracture limit based on stress-activated amorphization for alloys, relying exclusively on elemental properties[43] and tabulated energies[46] and lattice constants[45] from *ab initio* calculations.

***Dislocation slip stress and intrinsic ductility***

The next step in defining an intrinsic ductility is to define a limiting mechanism for slip (see **Figure 2a**). We begin by reviewing insightful work by Ogata et al.[52,53], which established a ductility prediction based on the stress for slip activation using the dislocation nucleation stress from the Frenkel model ($\tau_{Frenkel}$) normalized by the slip-system-dependent directional shear modulus ($G_{dir}$). They also explain how interstitial-charge density anisotropy accounts for deviations from the ideal sinusoidal generalized-stacking-fault energy (GSFE). They use DFT-calculated GSFE data for many metals to predict ductility, using the ratio of $\tau_{Frenkel}$ and surface energy, following the Rice criterion.

Two important considerations when defining a ductility parameter are the selection of (1) an energy *density* basis for slip activation and (2) an appropriate slip mechanism. The first point stems from the fact that the amorphization stress is an activation energy density, making it an intrinsic, bulk material property. In the same way that the amorphization activation energy density was shown to provide a lower energy mechanism for predicting the initiation of fracture compared to surface energy, a lower energy mechanism for slip in crystals exists. The Peierls barrier ($\tau_{Peierls}$), describes the activation energy for glide of *existing* dislocations. While it is insightful to compare ductility predictions based on both the dislocation glide and nucleation mechanisms, we focus our discussion of intrinsic ductility on comparisons with experimental data where plenty of dislocations are typically present before initiation of plastic deformation, making the glide (Peierls) barrier more appropriate for alloy design than the nucleation (Frenkel) barrier. The nucleation barrier is applicable when estimating the ductility of a perfect crystal.

The intrinsic ductility of metals, $\Delta$, is defined as the ratio of activation energy densities (indicated as bracketed stresses, $\langle\tau\rangle$) for the weakest slip mechanism, $\langle\tau_{slip}\rangle$, and for fracture initiation via amorphization, $\langle\tau_{amorph}\rangle$. Energy densities have units of pressure and are interpreted as deformation activation stresses.

$$\Delta = \frac{\langle\tau_{slip}\rangle}{\langle\tau_{amorph}\rangle} = \frac{\gamma_{usf} f_{bond}}{|\mathbf{b}|\langle\tau_{amorph}\rangle} \quad \textbf{Eq. 6}$$

Here, $\gamma_{usf}$ is the unstable stacking fault energy, with units of J/m$^2$, and |**b**| is the magnitude of the Burgers vector, with units of length, on the preferred slip system for each crystal structure. The parameter $f_{bond}$ is a unitless parameter that corrects for deviation from non-ideality in the shape of the GSFE curve, defined exclusively based on stiffness constants, and also shown to have a direct correlation to interstitial charge density; a detailed explanation is presented in the next section. The preferred slip systems used in this analysis are {110}<111> for BCC, {111}<112> for FCC, {0001}<11$\bar{2}$0> for basal HCP, and {10$\bar{1}$0}<11$\bar{2}$0> for prismatic HCP metals. For our ductility analysis, the limiting (maximum) Peierls barrier was selected on the basis that we are comparing the activation energy densities for two competing mechanisms at their maximum values. It is understood that the Peierls barrier must normally be reduced to account for the strain energy associated with strain from lattice disruption around the dislocation core if it is being used to predict experimental screw and edge dislocation activation energies and stresses. However, the same strain energy acts to depress the localized activation energy density needed to initiate amorphization, so a ratio of the two stresses inherently cancels dislocation core effect. This is analogous to the predictive accuracy that the amorphization model shows for both the strength of ultra-nanocrystalline metals[44], including their temperature dependence[43], and grain boundary energies[44] by factoring for localized structural and chemical disorder in reducing activation energy needed to initiate amorphization. When assessing ductility, taking the ratio of two activated processes that are linked by the shared activation volume has the simplifying effect that the strain effects must cancel.

***Interstitial Charge Redistribution Under Shear***

Ogata *et al.'s* analyses of the GSFE landscape for FCC, BCC, and HCP metals on the preferred (lowest potential energy) slip systems[52] showed that elements with interstitial electronic charge localization can deform with minimal stiffening as they are sheared (like Al and W), exhibiting "hinged-rod" bonding. Conversely, elements with spherical interstitial-charge gradients, (like Cu and Nb) exhibit "sphere-in-glue" bonding and significant stiffening during shear deformation on low-energy slip systems; this phenomenon is illustrated in **Figure 2b** using charge density cross-sections. They were successful in explaining why Al is disproportionately stronger than Cu, as a ratio of their shear moduli, which was attributable to dissimilar interstitial-charge distribution and validated in the context of their correspondingly large differences in unstable stacking fault energies (USFE), i.e., the barriers to slip. Ogata *et al.* also proposed an analytical expression for calculating $\tau_{Frenkel}$ based on directional shear modulus and the strain associated with peak stress, which was consistently limited to narrow (but differing) ranges for FCC, BCC, and HCP metals[53].

However, their model requires *a-priori* knowledge about what they termed "shearability" ($s_m$), which is the location on the GSFE curve corresponding to largest gradient, *i.e.*, highest stress. We define the parameter $f_{bond}$ (**Equation 7**) to describe the effect that charge distribution has on directional stiffness, and thus slip strength, by way of its effect on the GSFE. This approach is shown to be accurate for FCC, BCC, and HCP crystals, the latter requiring a differentiation between basal or prismatic slip. This generalization is an empirical observation linking interstitial-charge localization with directional stiffness in metals (and also crystal anisotropy – see supplemental), introducing a small but non-negligible correction to the activation stress for slip in metals. The directional shear moduli are defined based on stiffness constants for the conditions of simple shear ($G_{dir,u}$), where cell volume is unrelaxed giving $\varepsilon_{ij} = 0$ except for one shear strain component, and pure shear ($G_{dir,r}$), where cell volume is relaxed giving $\sigma_{ij} = 0$ except for one shear stress component. Their ratio indirectly captures the effect that shear deformation has on stiffness and strength, and thus charge density redistribution ($f_{bond}$).

$$f_{bond} = \frac{G_{dir,r}}{G_{dir,u}} = \frac{9C_{44}(C_{11} - C_{12})}{(C_{11} - C_{12} + 4C_{44})(C_{11} - C_{12} + C_{44})} \qquad \textbf{Eq. 7}$$

The introduction of $f_{bond}$ is important for establishing a more accurate correlation between slip stress based activation energy density ($\tau_{slip} \propto \gamma_{sfe}/|\mathbf{b}|$) and interstitial-charge density. Johnson *et al.*[45] showed that interstitial-charge density for metals and random concentrated mixtures (MPEAs and HEAs) can be derived via rule-of-mixtures of tabulated elemental values, as long as the correct charge density is used for each element as a function of its crystal structure in the alloy. A determination of the bond directionality for an alloy is possible using a similar approach. Shang *et al.*[47] provide *ab-initio* stiffness constants for elements (FCC, BCC, and HCP), which are used to establish a fundamental correlation between interstitial-charge density ($\rho_o$) and bond directionality, $f_{bond}$.

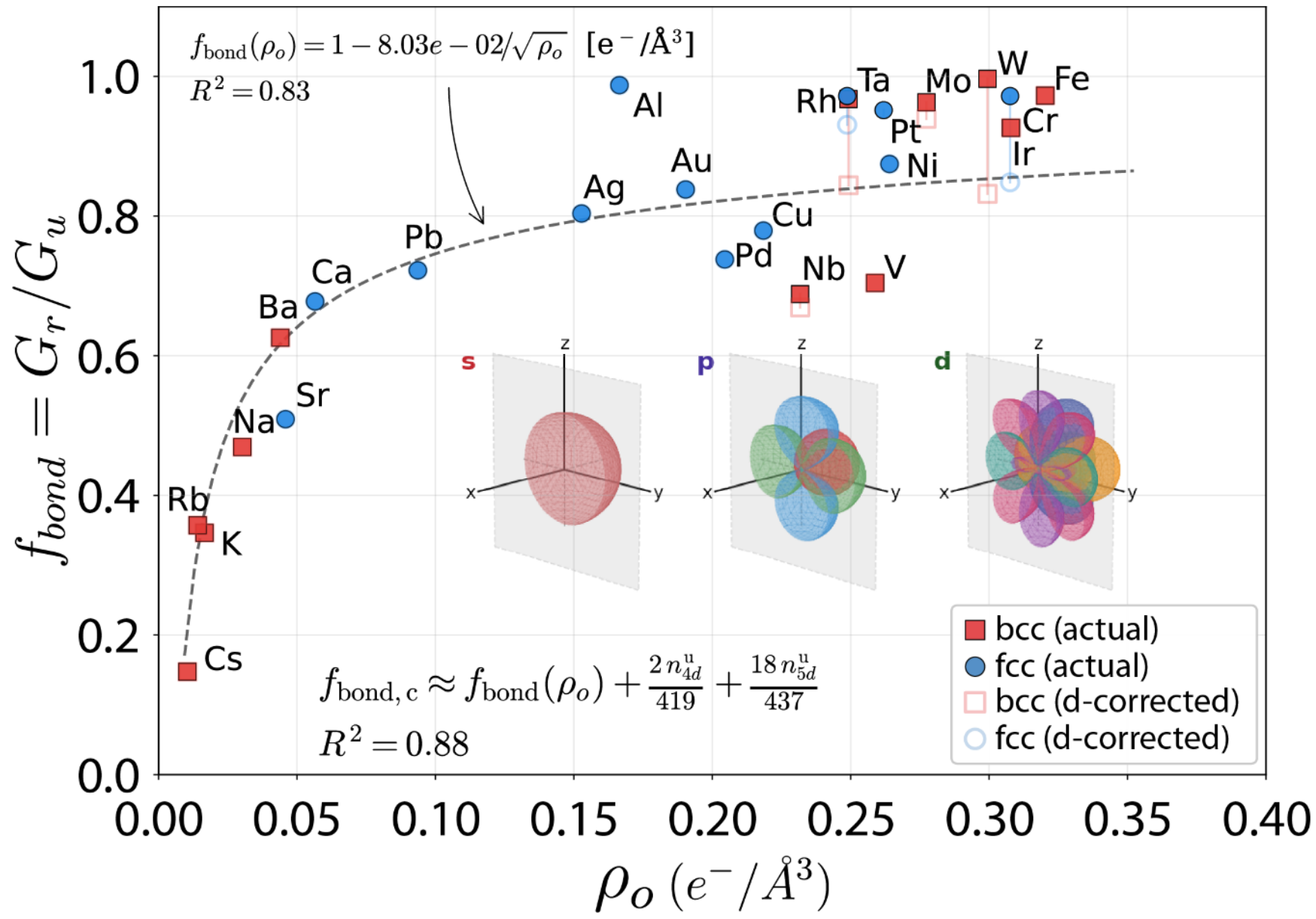

***Figure 3:** A comparison of bond directionality ($f_{bond}$) to interstitial charge density for FCC and BCC structures on their preferred slip systems. Note that basal HCP data are omitted as $f_{bond}$ = 1.*

An analysis of the relationship between charge density, $\rho_o$, and $f_{bond}$ is presented in **Figure 3**, giving the following empirical structure-property relationship for metals and alloys.

$$f_{bond}\left(\rho_o\right) = 1 - \frac{0.08}{\sqrt{\rho_o}} \left(\frac{e^-}{\AA^3}\right) \qquad \textbf{Eq. 8}$$

Deviations from the baseline asymptotic power law (residual error) linking activation stresses for slip to $\rho_o$ are hypothesized to be linked to valence electron count and band filling, *i.e.*, Fermi energy occupancy for the density-of-states at T = 0 K. Factors that affect bond directionality are likely to include spin-orbit coupling (*d*-band splitting), relativistic effects (*s*-contraction and *d*-expansion, most prevalent in heavier 5*d* transition metals like Au), and *s*-*d* orbit hybridization (*e.g.*, Cu and Pd). As a first-order approximation, the same plot shows an overlaid correction based on unpaired electrons by orbital, with a single weighting factor for *s*-, *p*-, 3*d*-, 4*d*-, and 5*d*-orbitals. This is meant to show that the residual errors from the mean power law can be largely explained by asymmetry in orbital occupation. Additional DFT analysis is needed establish direct links between interstitial-charge density asymmetry, band-filling, and bond-directionality, possibly via the electron localization function (ELF)[54–56]. The present work suggests that it may be possible to develop an analytical expression that directly links band-filling and bond-directionality, providing a more accurate prediction of the intrinsic strength and ductility of alloys. In lieu of this refinement, we adopt the asymptotic power law correlating $f_{bond}$ and $\rho_o$ which is sufficiently accurate at describing bond-directionality effects on slip activation, $\langle\tau_{slip}\rangle$, enabling ductility predictions for pure metals and concentrated solid solutions.

***The Physical Link between Charge Density and Activation Energy Densities***

The interstitial charge density, $\rho_{eff}$, and bond directionality parameter, $f_{bond}$, have no interdependence ($R^2$ ~ 0.015), establishing that these two parameters are capturing different physics. Charge density captures bond strength and the bond directionality parameter captures the *change* in bond strength due to strain, a measure of bond stiffness. **Figures 1b** shows that these are structure-property relationships for metals and alloys, enabling predictions of $\langle\tau_{amorph}\rangle$ (**Eq. 9**) and $\gamma_{usf}/|\mathbf{b}|$ (**Eq. 10**) from interstitial charge densities corrected for crystal packing fraction. The latter correction indicates that the universal behavior is governed by the interstitial charge participating in bonding normalized by the atomic (Wigner-Seitz) volume, rather than the interstitial volume.

$$\langle\tau_{amorph}\rangle = 357\,\rho_{eff}^{1.85} \qquad (R^2 = 0.89) \qquad \textbf{Eq. 9}$$

$$\frac{\gamma_{usf}}{|\mathbf{b}|} = 3840\,\rho_{eff}^{2.77} \qquad (R^2 = 0.91) \qquad \textbf{Eq. 10}$$

The ratio of these relationships gives the following correlation between interstitial charge density, $\rho_{eff}$, bond stiffness, $f_{bond}$, and intrinsic ductility, $\Delta$.

$$\Delta = 10.8\,f_{bond}\rho_{eff}^{0.92} \qquad \textbf{Eq. 11}$$

**Equations 9 and 10** are empirical scaling laws, linking physical properties of materials (interstitial charge densities) to their response to external stimuli (characteristic stress scales for slip nucleation and amorphization). The power law exponents and prefactors have physical origin. The Thomas-Fermi (T-F) model for a free electron gas, which is a simplification for the Kohn-Sham equations that form the basis of these DFT calculations, provides insight. By ignoring exchange-correlation, ionic interactions with nuclei, and band structure, which require numerical solutions via models like Kohn-Sham, the Thomas-Fermi model gives an analytical first-principles correlation between bulk modulus and charge density from the kinetic energy of an electron gas, $B_{T\text{-}F} = 3895\rho_{eff}^{5/3}$. The exponent 5/3 (~1.67) slightly underpredicts ($\approx$ 10%) the exponent 1.85 in **Eq. 9**, which was determined using Kohn-Sham and thus captures the added effects of band structure, ionic interactions, and exchange-correlation. It is reasonable that the additional physics results in slightly steeper scaling as a function of charge density.

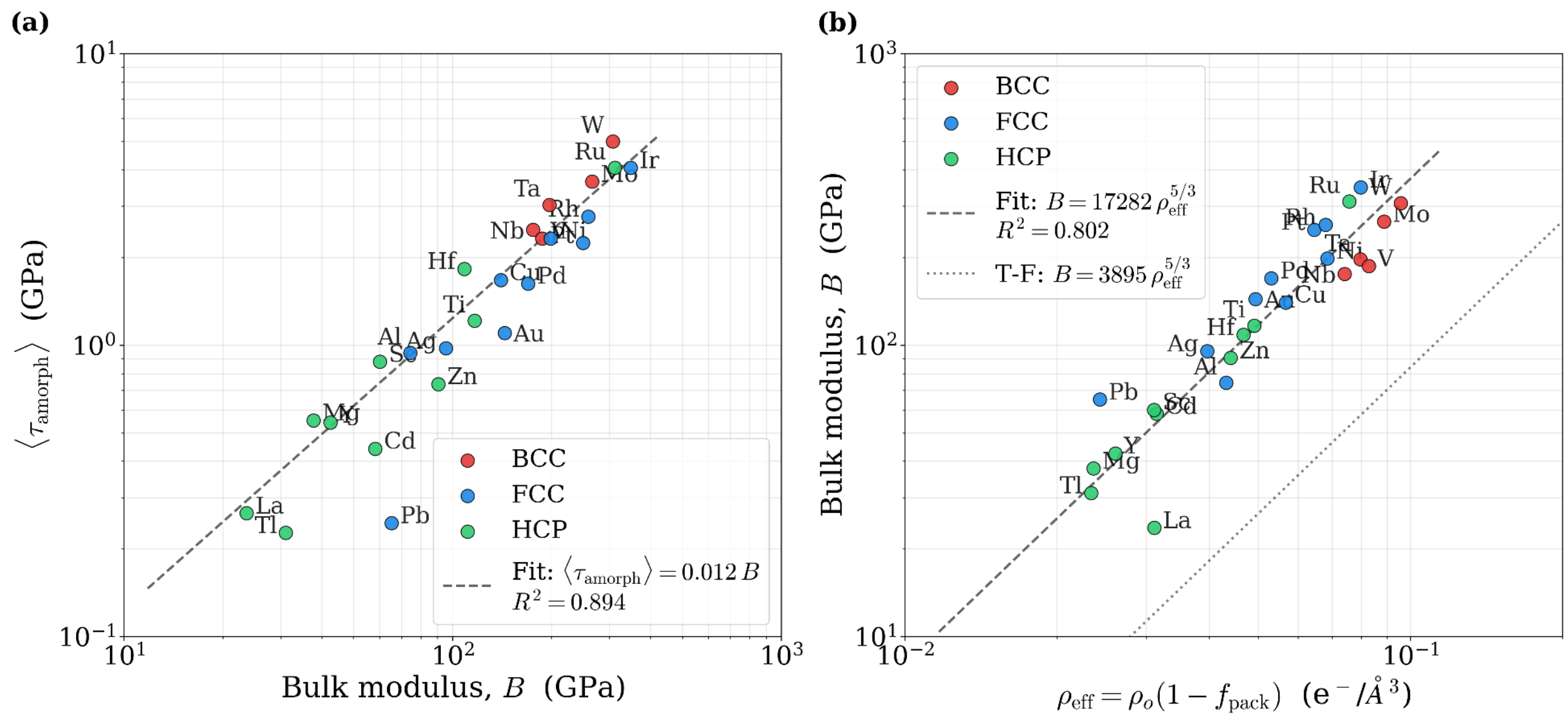

***Figure 4:*** *Relationships between (a) amorphization energy density and DFT bulk moduli [47] (T = 0 K) and DFT bulk moduli to effective interstitial charge density.*

The empirical relationship between amorphization energy density (bond strength) and bulk modulus, the latter from DFT [47] at T = 0 K, is shown in **Figure 4a** to be $\langle \tau_{amorph} \rangle \approx 0.012B$. The Thomas-Fermi prefactor 3895 GPa/($e^-$/Å$^3$) is exclusively geometrical in origin, but, again, based only on charge repulsion in a free electron gas. **Figure 4b** shows that the additional physics in the Kohn-Sham model requires a scaling factor, 17282/3895 ≈ 4.44, even though the 5/3 exponent remains reasonably accurate. Combining these equations gives, $\langle \tau_{amorph} \rangle \approx 0.012\left(17282\rho_{eff}^{5/3}\right) = 207\rho_{eff}^{5/3}$, which is comparable to **Eq. 9**. This means that **Eq. 9** is, as expected, a more physically accurate scaling law for the bond strength on the basis of interstitial charge density. As there is no analytical expression for shear modulus, it is not possible to provide a similar comparison for **Eq. 10**, but it is clear that the prefactor and exponent are describing the fundamental relationship between interstitial charge density and normalized unstable stacking fault energy via a scaling factor that represents the slip-system dependent shear stiffness and an intrinsically steeper correlation with charge density. A systematic investigation of hydrostatic and shear strain on interstitial charge-density is needed to properly decouple the physics of band-structure, charge density, and charge redistribution under shear, which should clarify the mechanistic origin of **Eqs. 9 and 10**. However, the present approximation appears to be useful as an improvement over existing ductility prediction models.

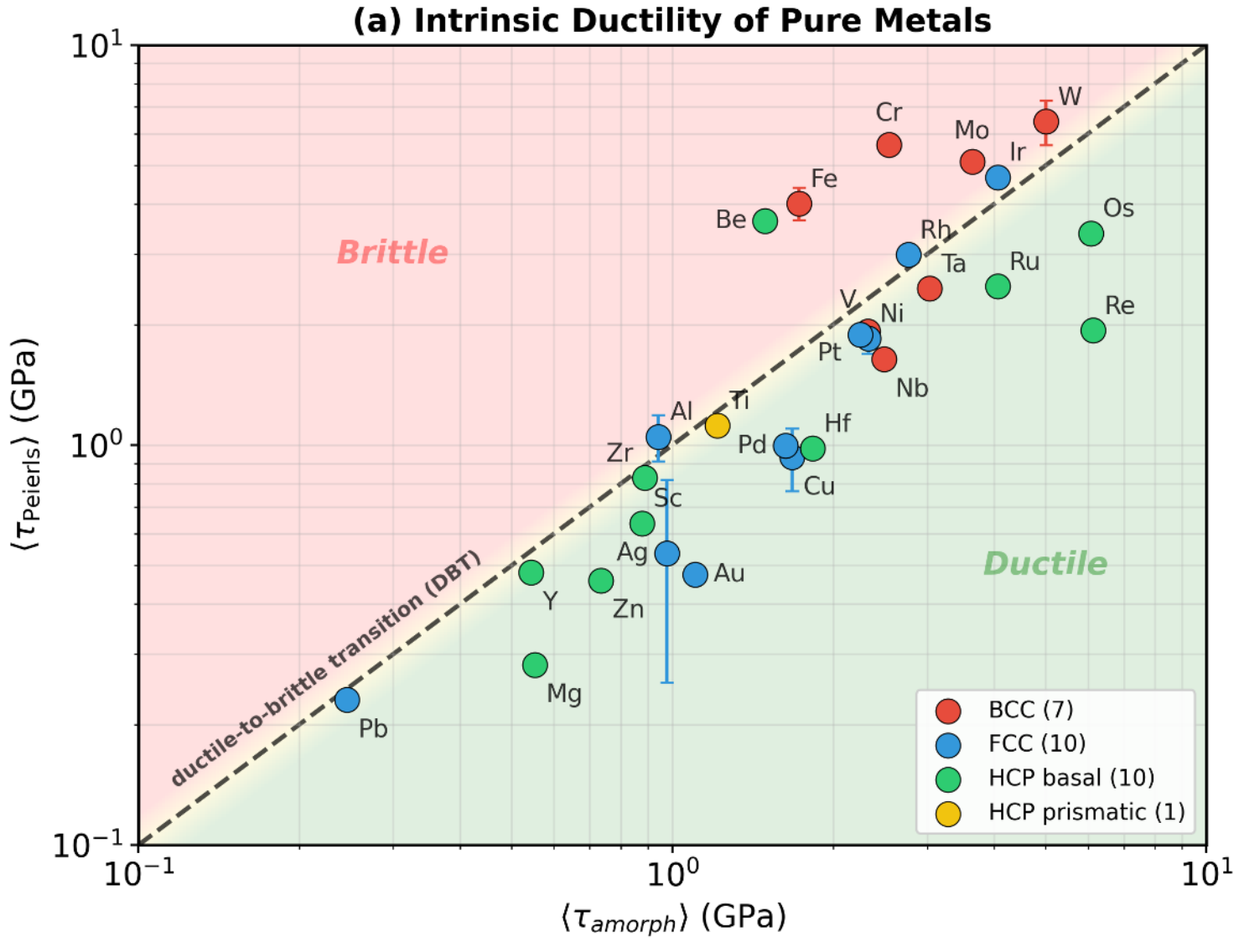

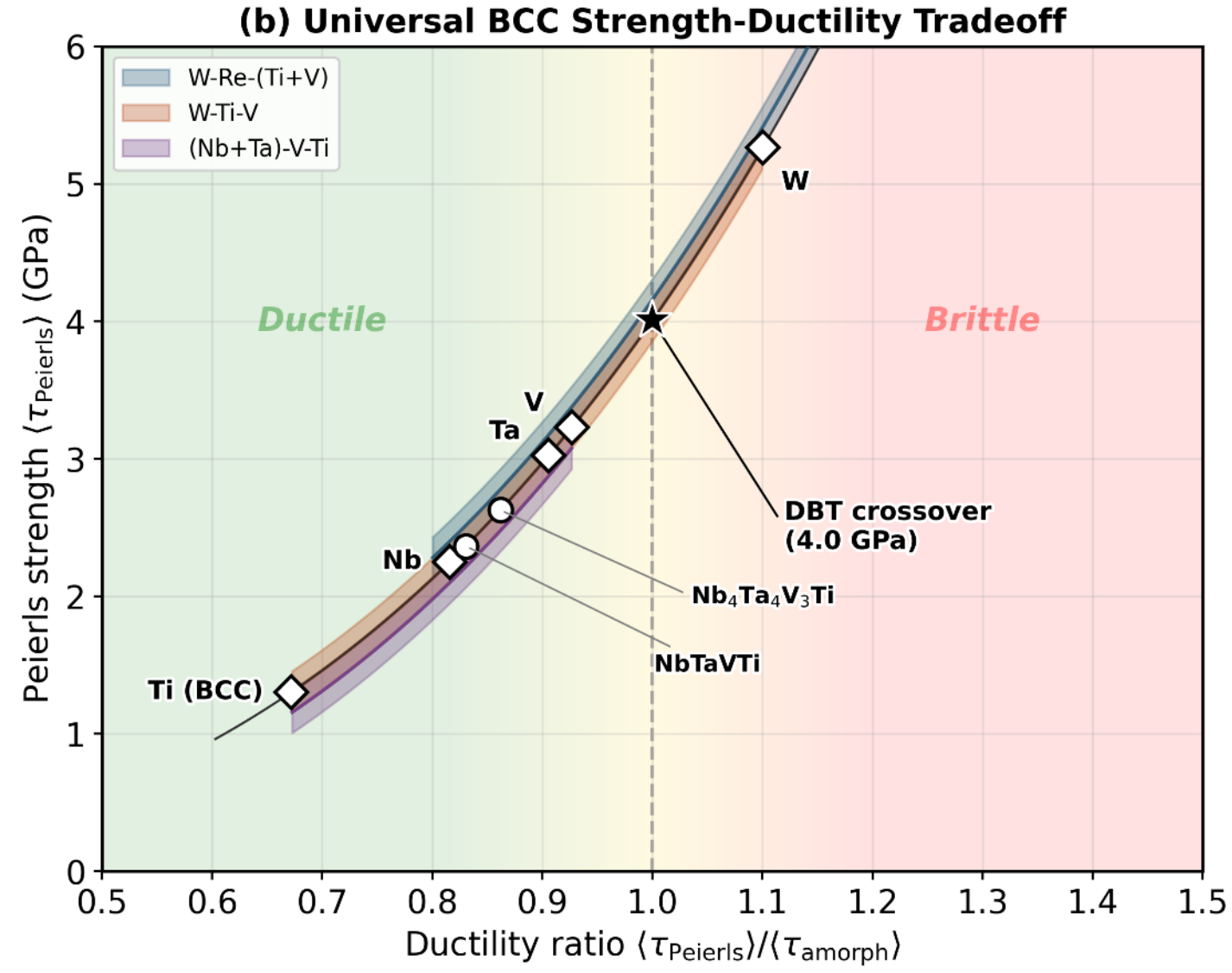

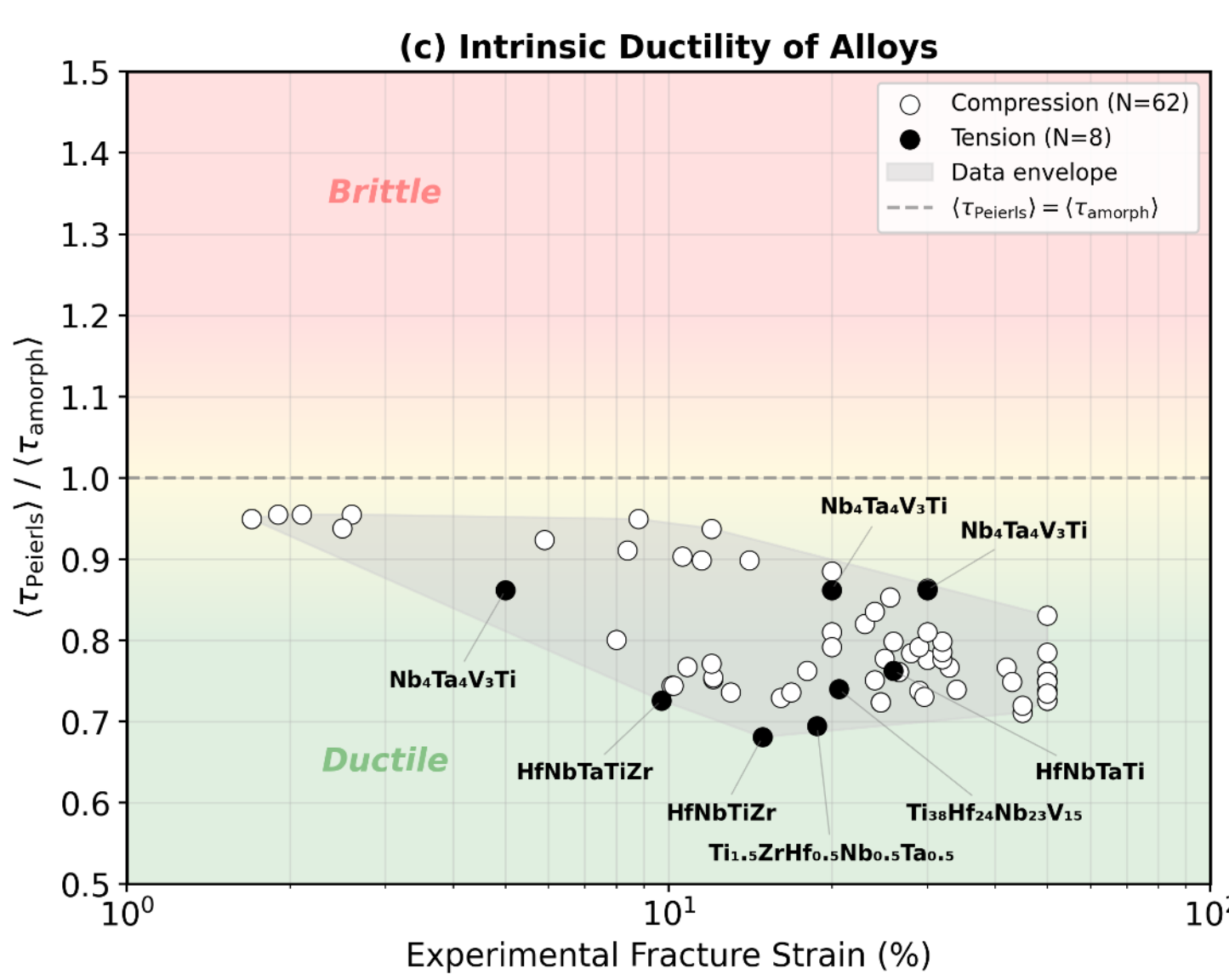

***Figure 5:*** *(a) Correlation between energy-density based activation stresses for slip and amorphization,* $\langle \tau_{amorph} \rangle$ ***(Eq. 1-2)*** *and* $\langle \tau_{Frenkel} \rangle$ ***(Eq.*** *7), showing the regimes of intrinsic ductility or brittleness for elements; error bars correspond to the variability in DFT-based calculations of USFE*[57];

*(b) The relationship between dislocation-slip strength **(Eq.** 7) and intrinsic ductility **(Eq. 13)** for exemplar multi-principal-element alloys Nb-Ta-V-Ti, W-Ti-V, and W-Re-Ti-V, indicating that the relationship between these properties and interstitial-charge density reveal a universal curve.*

*(c) Comparison of the predicted intrinsic ductility of multiple single-phase MPEAs vs their experimentally measured tensile and compressive fracture strains at room temperature*[27,37,41,58–64].

***Alloy Predictions and Validation***

Validation of the proposed framework is possible by analyzing available data for elements and alloys. As was shown by Johnson *et al.*[45], the interstitial charge density of concentrated, random metal solid solutions can be determined with high accuracy using a rule-of-mixtures of the DFT-tabulated elemental interstitial charge densities for lattice-matched configurations; *i.e.*, a BCC alloy of many metals has an interstitial charge density, $\rho_{o,alloy} = \sum x_i \rho_{i,bcc}$. Combining **Eq. 6-10** with the tabulated binary interaction energies from Chen *et al.*[46], to determine mixing enthalpy for alloys, enables the direct prediction of intrinsic ductility for pure metals (**Figure 5a**) and alloys. A plot of the change in strength and ductility as a function of composition for a few stable single-phase BCC alloy systems, shown in **Figure 5b**, reveals a universal trade-off curve based on their generalized response via interstitial-charge density. A comparison of calculated intrinsic ductility and experimental room-temperature fracture strain in tension and compression is presented in **Figure 5c** for a range of refractory (BCC) concentrated alloys. While the important effect that uniaxial compressive stress can have on delaying fracture in otherwise brittle materials by closing cracks as they initiate, contrasting with uniaxial tension, the predicted ductile-to-brittle transition shows good agreement with available experimental data.

Using the the same analytical expressions, it is also possible to rapidly generate phase diagrams for refractory (BCC) MPEAs showing their intrinsic strength and ductility, and an estimate of phase stability based on the criterion that single-phase mixtures will be stable when -10 kJ/mol < $H_{mix}$ < +5 kJ/mol. When mixing enthalpy exceeds 0 kJ/mol, phase separation becomes energetically favorable, once configurational entropy is overcome. For large negative mixing enthalpy, the formation of ordered compounds (intermetallic phases) becomes likelier. **Figure 6** shows predictions of slip activation stress, intrinsic ductility, and phase stability in the form of pseudoternary phase diagrams for two exemplar MPEAs. The Nb-Ta-V-Ti system was selected as it has been shown to exhibit high strength and room-temperature ductility in uniaxial tension[37], and the W-Ti-V alloy system[65], which is of interest for plasma containment in next-generation fusion reactors. The choice of Nb+Ta as a paired set in the pseudoternary plots is based on commonality in *d*-band filling and high/low spin-state fitting.

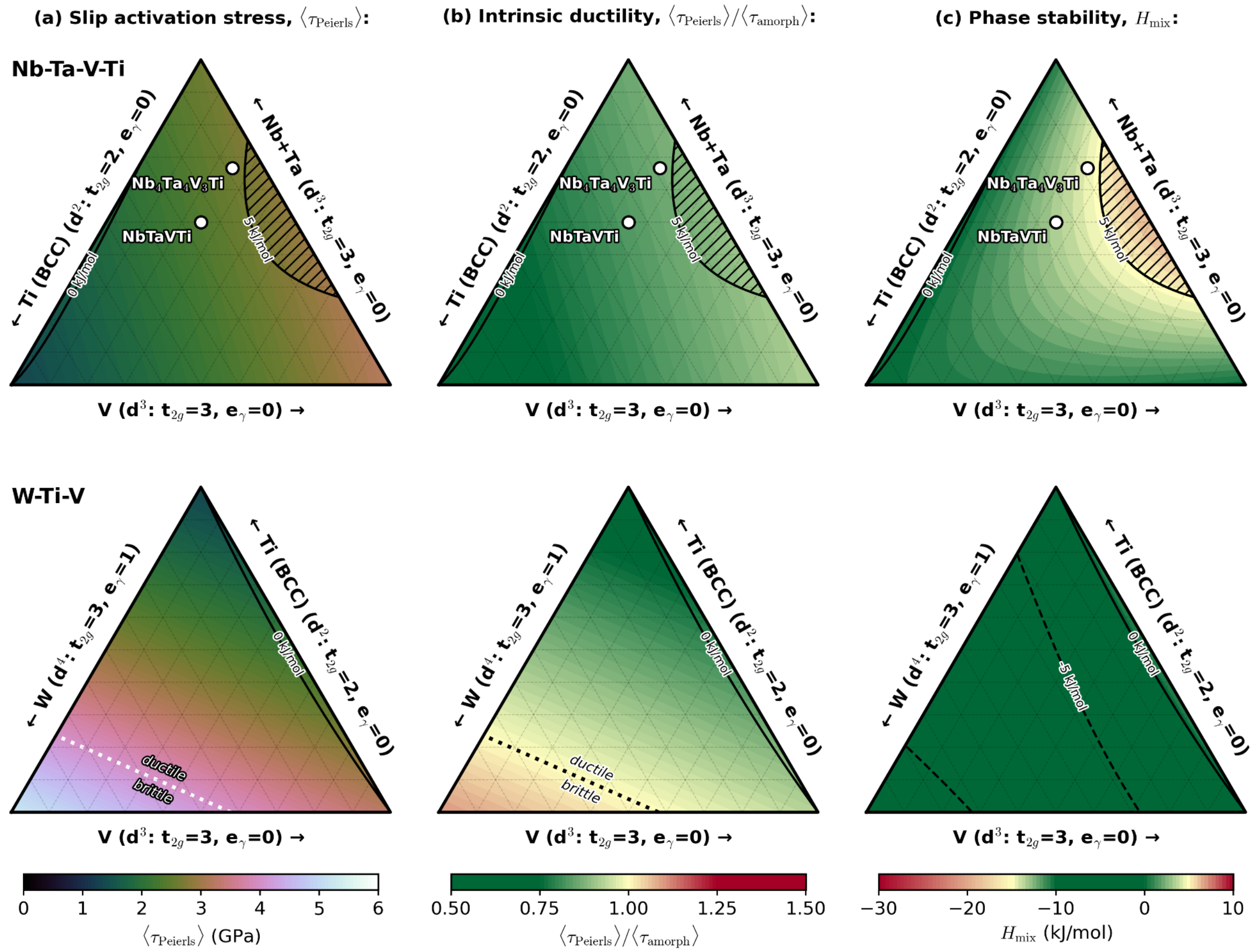

***Figure 6:*** *(a) Analytically predicted slip-activation intrinsic strength, (b) intrinsic ductility, and (c) phase stability based on formation enthalpies from Chen et al.[46], using binary interaction energies from Chen et al.[46], for Nb-Ta-V-Ti and W-Ti-V, where the pairing of Nb+Ta to enable pseudoternary plotting was selected based on similarly in band structure (similar d-band electron and high/low spin state filling). Nb-Ta-V-Ti plots include equiatomic NbTaTiV and* $Nb_4Ta_4V_3Ti$*, both shown to exhibit room-temperature ductility and varying degrees of tensile strength*[30,37,58].

**A Framework for Estimating Ductile-to-Brittle Transition Temperature (T > 0 K)**

Work hardening in BCC metals like W and Mo is associated with an increase in dislocation density ($\rho_{disl}$) and reduction in grain size ($d_{grain}$). Dislocations tend to aggregate into dislocation cells, which along with grain boundaries become effective barriers for dislocation slip. The effect of dislocation density on the flow or yield strength is shown for polycrystalline W in **Figure 7a** (see supplemental for Mo). Tungsten was shown to be intrinsically brittle (**Figure 5a**); however, it is also known to exhibit ductility at temperatures as low as room-temperature via extreme grain refinement[66]. The ductile-to-brittle transition temperature ($T_{DBT}$) is the result of complex crossovers in mechanisms of deformation that are a function of strain rate, purity, and defect density.

As a demonstration of how the present framework can be used to explain these complex phenomena, we explore the problem of competition between slip and amorphization by modifying the idealized energy-density stress equations to account for the effect of localized strain fields, specifically those generated when dislocations aggregate and interact with slip barriers including grain boundaries and dislocation cell walls. Similar to how Rice[21] calculated the strain effect near a crack tip based on an energy release rate, where the transition between ductile and brittle behavior is set by the stress required to propagate a crack, we apply the same rationale to determine the critical stress, $\tau_{crit}$, required to nucleate an amorphous interface. The nucleation event is analogous to nucleation of a high-energy grain boundary in the vicinity of a stress concentration. For edge dislocations[22], the energy release rate is treated as a mode II shear-loaded crack and expressed as a function of Poisson ratio, $\nu$, shear modulus, $\mu$, and stress concentration factor, $K_{II}$, where the latter can be reduced to an applied stress, $\tau_{app}$, and the characteristic length between slip barriers, $L_b$. For screw dislocations[22], the energy release rate is treated as a mode III shear-loaded crack.

$$G_{edge} = \frac{(1-\nu)}{2\mu} K_{II}^2 = \frac{(1-\nu)}{2\mu} \left(\tau_{applied}\sqrt{\pi L_b}\right)^2 \qquad \textbf{Eq. 12}$$

$$G_{screw} = \frac{1}{2\mu} K_{III}^2 = \frac{1}{2\mu} \left(\tau_{applied}\sqrt{\pi L_b}\right)^2 \qquad \textbf{Eq. 13}$$

The J-integral applied to a crack tip[21] also shows that there is an equivalence between energy release rate and the energy associated with forming new crack surface, $G = 2E_{surf}$. Applying a similar equivalence based on amorphization rather than cleavage, $G = \langle \tau_{amorph} \rangle \cdot 2|b|$, to **Equations 12** and **13** results in reduced expressions for the applied critical stress as a function of $\langle \tau_{amorph} \rangle$, where the localized strain energy is sufficient to nucleate amorphization where one or more dislocations pile-up at a barrier. Experimental observations of deformed refractory BCC metals reveal mixed edge and screw dislocation populations at $T \geq T_{DBT}$, and long straight screw dislocations on $\{110\}$ planes at $T < T_{DBT}$[67,68]. For W, we used **Equation 14**, but note that given the range of Poisson ratios for refractory metals is about 0.3, the stress difference is only ~ 15%.

$$\tau_{crit,edge} = \sqrt{\frac{4\mu|\boldsymbol{b}| \langle \tau_{amorph} \rangle}{\pi (1-\nu) L_b}} \qquad \textbf{Eq. 14}$$

$$\tau_{crit,screw} = \sqrt{\frac{4\mu|\boldsymbol{b}| \langle \tau_{amorph} \rangle}{\pi L_b}} \qquad \textbf{Eq. 15}$$

The remaining unknown is the characteristic length, $L_b$, which represents the distance between barriers for dislocation pile-up; this length also establishes the size of the volume over which the strain field associated with dislocation pileups is dissipated. We define these for grain boundaries and dislocation cell walls, as these are shown to be remarkably accurate simplifications for which length scales are readily available, limiting barrier length to the smaller of $L_b = d_{grain}/2$ and $d_{cell}/2$. This also enables direct correlations with dislocation areal density, $\rho_{disl}$, where for refractory metals[69], $d_{cell} \approx K\rho^{-0.5}$, where $K$ values range from 20-30 for refractory metals[70].

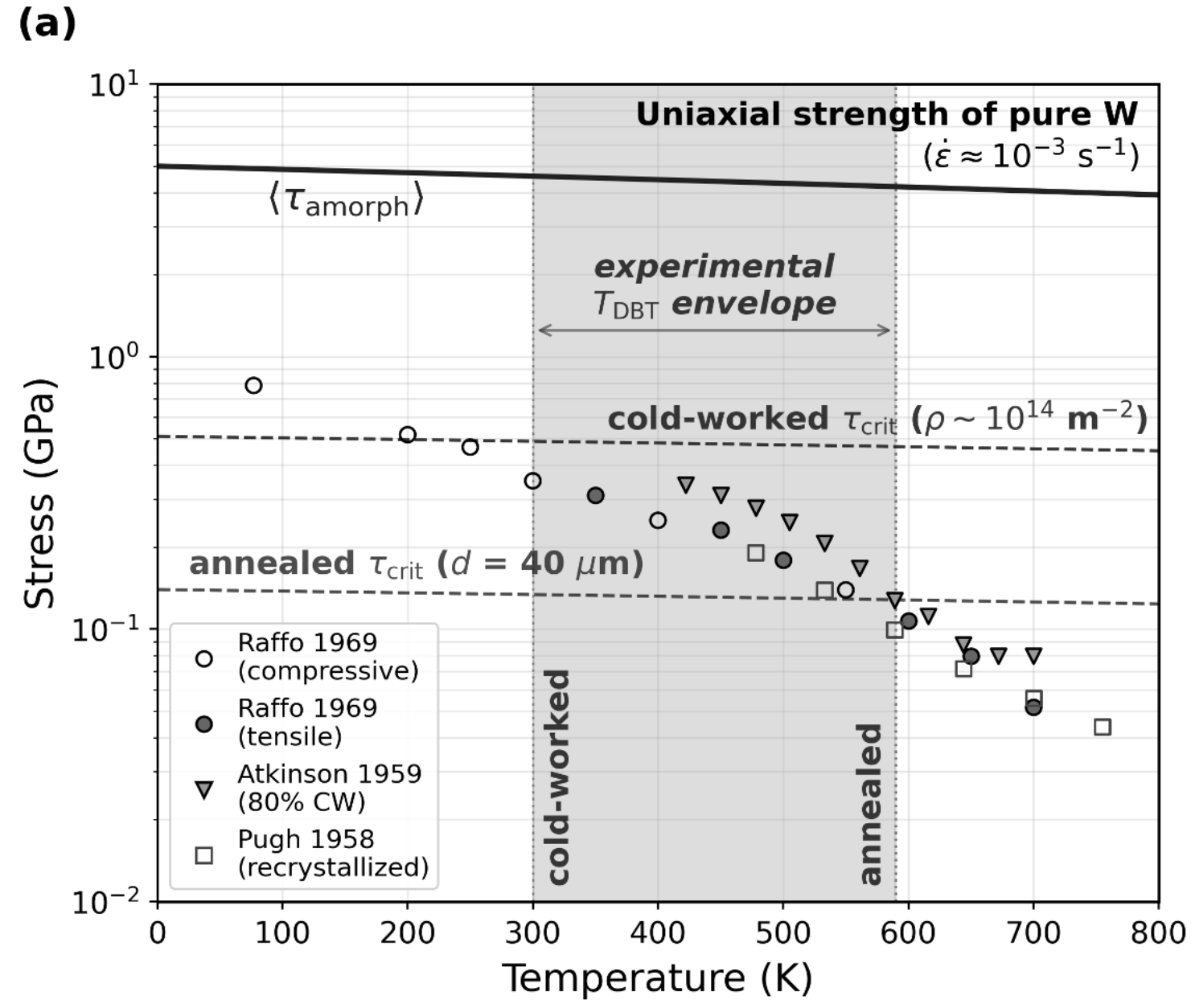

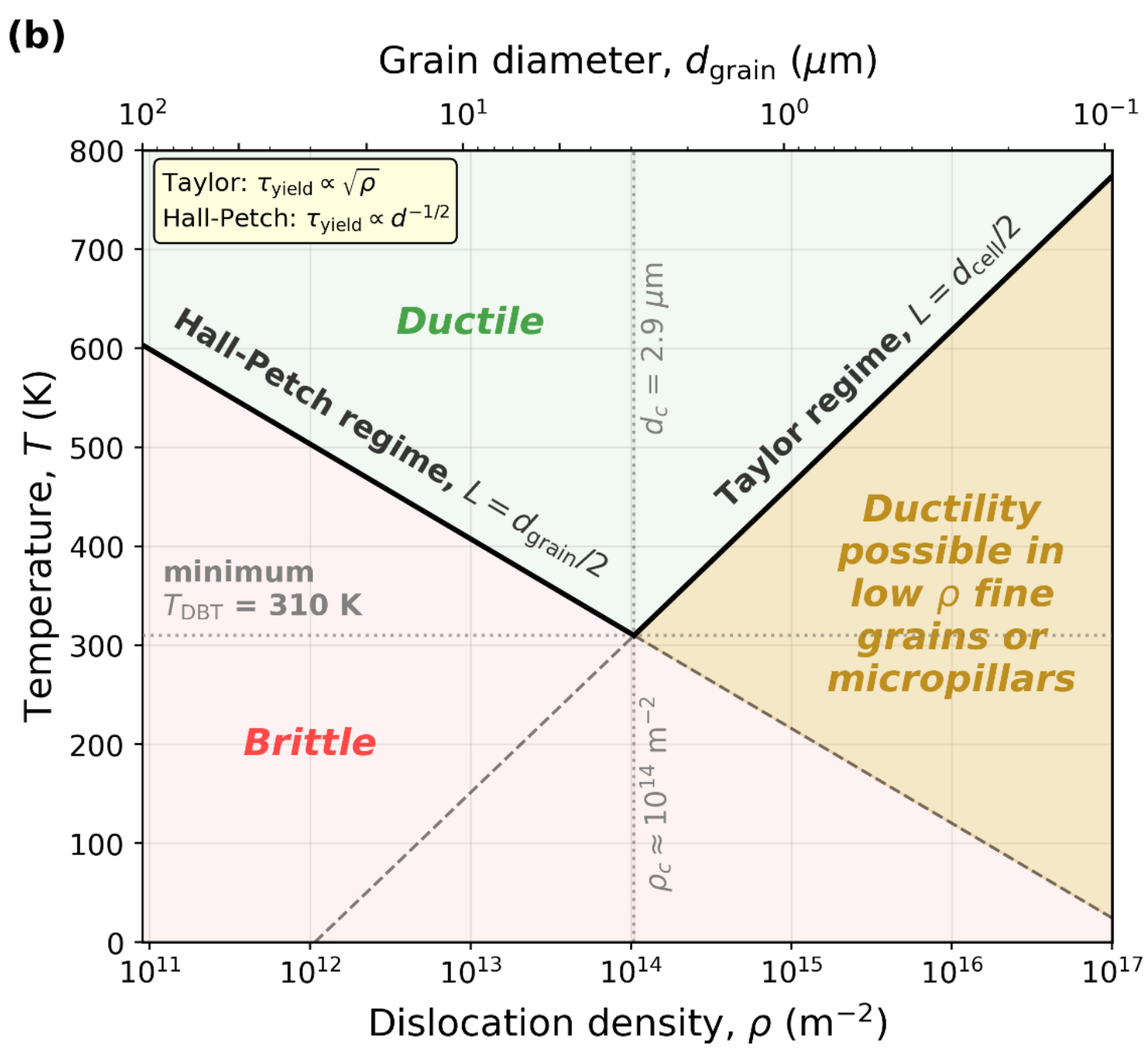

***Figure 7:*** *(a) Experimental tensile yield strength data for high-purity polycrystalline tungsten at quasi-static strain rates from multiple sources[71–73], a shaded envelope showing the range of reported ductile-to-brittle transition temperatures ($T_{DBT}$), and dashed lines corresponding to predicted DBT transition stresses, $\tau_{crit}$ (**Eq. 14**), for the reported grain sizes and dislocation densities in annealed and cold-worked conditions. The intersection of the predicted DBT ($\tau_{crit}$) and experimentally-determined yield strengths at the $T_{DBT}$ for both annealed and cold-worked pure tungsten show good agreement.*

*(b) This plot shows the correlation between dislocation density (lower axis), grain size (upper axis), and temperature (left axis) on $\tau_{crit}$, where $T_{DBT}$ is predicted as the locus of points that make up the two solid lines (**Eq. 14**), i.e., $\tau_{crit}$ for $L_b$ = min($d_{cell}$, $d_{grain}$).*

**Figure 7a** presents experimental temperature-dependent, quasi-static polycrystalline yield strength data from multiple sources for high-purity W and overlaid predictions of $\tau_{crit}$. The ductile-to-brittle transition temperature is expected where $\tau_{crit}$ intercepts yield strength. The shaded envelope corresponds to the reported experimental $T_{DBT}$ limits for these bulk polycrystalline microstructures. **Figure 7b** further explores the implications of **Equations 14-15**, dislocation density ($\rho_{disl}$) and grain size in intrinsically brittle metals like W and Mo. At constant temperature and strain rate, $\tau_{crit}$ is proportional to $\rho_{disl}^{0.25}$, implying that the ductile-to-brittle transition temperature, $T_{DBT} \propto \rho^{0.25}$. When dislocation density is very small, as with annealed material, $\tau_{crit}$ and thus $T_{DBT}$ are a function of grain size, and for single crystals, as has been shown via micropillar compression, a function of sample volume size[74]. The unifying feature is that the volume of the smallest barrier separation distance (grain, dislocation cell, or sample size) defines the strain field generated by dislocation pile-ups interacting with the respective barriers (grain boundaries, cell walls, or sample surfaces). **Figure 7b** identifies the possibility of extending the $T_{DBT}$ to values asymptotically approaching 0 K if nanocrystalline samples can be manufactured with few intracrystalline defects, *e.g.*, synthesis of micro- and nano-pillars from single crystals.

## Conclusions

We establish direct analytical expressions based on empirical relationships between interstitial-charge density and limiting intrinsic strength mechanisms for metals and alloys. The assumptions are few (and often historical), and the errors associated with them appear to be sufficiently small to justify the use of this approach for rapid quantitative alloy discovery. There is evidence that additional work may elucidate improvements in the accuracy of these methods, such as through the examination of the links between band-filling and charge heterogeneity (bond-directionality). Additionally, these results lay a foundation for the possible prediction of ductile-to-brittle transition temperature shifts associated with interstitial impurities. Systematic *ab-initio* calculations of charge-density changes due to non-metal interstitial inclusions may provide rules for calculating localized changes in strength, and thus ductility.

## Methods

### *Density functional theory calculations*

All-electron charge densities for 41 metallic elements across three crystal structures (BCC, FCC, HCP) were computed using the projector augmented-wave (PAW) method[75] as implemented in Quantum ESPRESSO v7.5[76,77]. The Perdew-Burke-Ernzerhof (PBE) generalized gradient approximation[78] was employed for the exchange-correlation functional, with PAW datasets from the PSlibrary 1.0.0[79]. Kinetic energy cutoffs of 60 Ry (wavefunction) and 480 Ry (charge density) were used for transition metals, reduced to 40 Ry and 320 Ry for *s*- and *p*-block metals. Brillouin zone integrations used 12 x 12 x 12 Monkhorst-Pack *k*-point meshes[80] for cubic structures and 12 x 12 x 8 for hexagonal structures, with Fermi-Dirac smearing of 0.007 Ry. Self-consistency was converged to $10^{-10}$ Ry in total energy.

Lattice constants were taken from the DFT-PBE relaxed values of Shang *et al.*[47], which are also those used by Johnson *et al.*[45] in their interstitial electron count (IEC) tabulation. For HCP metals, the *c/a* ratios from Johnson *et al.* were adopted. This ensures direct comparability between our charge density calculations and the IEC values used throughout this work for elements and alloys.

***Interstitial charge density extraction***

The interstitial charge density, $\rho_o$, was extracted from the all-electron charge density on a $200^3$ real-space grid via the post-processing tool `pp.x`. Adapting the concept by Johnson *et al.*[45] to a real-space extraction, we define $\rho_o$ as the average charge density within the largest inscribed sphere that fits in the interstitial region of the crystal, *i.e.*, the sphere centered at the bond midpoint between nearest-neighbor atoms, with radius equal to the interstitial radius $r_{\mathrm{IS}} = (d_{\mathrm{NN}} - 2r_{\mathrm{atom}})/2$, where $d_{\mathrm{NN}}$ is the nearest-neighbor distance and $r_{\mathrm{atom}}$ is the metallic radius for coordination number 12. The interstitial electron count is then $\mathrm{Q}_o = \rho_o \cdot \Omega_o$, where $\Omega_o$ is the interstitial volume. This procedure was applied to all 122 element-structure combinations (41 elements x 3 structures), yielding PAW-PBE reference values of $\rho_o$ with a mean absolute deviation of 3.9% from Johnson *et al.*'s KKR-ASA values[45].

**Acknowledgments**

The authors thank Michael J. Abere, Hailong Huang, and Prashant Singh for their helpful conversations and comments on this work. NA was supported by a grant from The Scientific Review, a 501(c)(3) organization, and is grateful for financial, logistical, and technical support from Lynzie R. Rowland. MRJ was supported by the Laboratory Directed Research and Development (LDRD) program at Sandia National Laboratories. DDJ was supported by the U.S. Department of Energy (DOE) ARPA-E CHADWICK program. This work was produced in part by Sandia National Laboratories, a multiprogram laboratory managed and operated by National Technology and Engineering Solutions of Sandia, LLC, a wholly owned subsidiary of Honeywell International, Inc., for the U.S. Department of Energy's National Nuclear Security Administration under contract DE-NA0003525, and in part by Iowa State University under Contract No. DE-AC02-07CH11358 with the U.S. Department of Energy. This paper describes objective technical results and analysis. Any subjective views or opinions that might be expressed in the paper do not necessarily represent the views of the U.S. Department of Energy or the United States Government.

**Author contribution statements**

MRJ contributed to theory development, computational work, analysis and interpretation, writing, and editing. DDJ contributed to conception, theory development, interpretation, and editing. NA contributed to conception, theory development, analysis and interpretation, writing, editing, and project management.

**Competing Interest Statement:** The authors declare no competing interests.

**Data Availability:** All relevant data provided as supplemental material, including HDF5 and ODT formats of all tabulated parameters.

**Code Availability:** No unique codes were developed for the purpose of this work. Computational output results are provided as Supplementary Material.